\documentclass[fleqn,usenatbib]{mnras}

\usepackage{newtxtext,newtxmath}
\usepackage[linewidth=1pt]{mdframed}
\usepackage{color}

\usepackage[T1]{fontenc}
\usepackage{CJKutf8}
\DeclareRobustCommand{\VAN}[3]{#2}
\let\VANthebibliography\thebibliography
\def\thebibliography{\DeclareRobustCommand{\VAN}[3]{##3}\VANthebibliography}

\usepackage{graphicx}	
\usepackage[export]{adjustbox}
\usepackage{amsmath}	
\usepackage{float}
\usepackage{subfigure}

\title[Superflare GWAC\,220525A]{A huge-amplitude white-light superflare on a L0 brown dwarf discovered by GWAC survey}

\author[Xin et al.]{
Li-Ping Xin$^{1}$\thanks{E-mail: xlp@nao.cas.cn},
Hua-li Li$^{1}$,
Jing Wang$^{2,1}$\thanks{E-mail: wj@nao.cas.cn}, 
Xu-Hui Han$^{1}$,
Hong-Bo Cai$^{1}$,
Xin-Bo Huang$^{2}$,
Jia-Xin Cao$^{2}$,
\newauthor
Yi-Nan Zhu$^{3}$,
Xiang-Gao Wang$^{2}$,
Guang-Wei Li$^{1}$,
Bin Ren$^{2,1,4}$,
Cheng Gao$^{2,1,4}$,
Da Song$^{2,1,4}$,
Lei Huang$^{1}$,
\newauthor
Xiao-Meng Lu$^{1}$,
Jian-Ying Bai$^{1}$,
Yu-Lei Qiu$^{1}$,
En-Wei Liang$^{2}$,
Zi-Gao Dai$^{5,6}$,
Xiang-Yu Wang$^{6,7}$,
\newauthor
Chao Wu$^{1}$,
Jing-Song Deng$^{1,4}$,
Yuan-Gui Yang$^{8}$,
Jian-Yan Wei$^{1,4}$
\\
$^{1}$CAS Key Laboratory of Space Astronomy and Technology, National Astronomical Observatories, Chinese Academy of Sciences, Beijing 100101, China.\\
$^{2}$Guangxi Key Laboratory for Relativistic Astrophysics, School of Physical Science and Technology, Guangxi University, Nanning 530004, China.\\
$^{3}$ Key Laboratory of Optical Astronomy, National Astronomical Observatories, Chinese Academy of Sciences, Beijing 100012, Peopleʼs Republic of China.\\
$^{4}$School of Astronomy and Space Science, University of Chinese Academy of Sciences, Beijing, China.\\
$^{5}$Department of Astronomy, University of Science and Technology of China, Hefei 230026, Peopleʼs Republic of China.\\
$^{6}$School of Astronomy and Space Science, Nanjing University, Nanjing 210093, China.\\ 
$^{7}$Key Laboratory of Modern Astronomy and Astrophysics (Nanjing University), Ministry of Education, Nanjing 210093, China.\\
$^{8}$School of Physics and Electronic Information, Huaibei Normal University, Huaibei 235000, China.
}

\date{Accepted XXX. Received YYY; in original form ZZZ}

\pubyear{}

\begin{document}
\begin{CJK}{UTF8}{gbsn}
\label{firstpage}
\pagerange{\pageref{firstpage}--\pageref{lastpage}}
\maketitle

\begin{abstract}
White-light superflares from ultra cool stars are thought to be resulted from magnetic reconnection, but the magnetic dynamics in a fully convective star is not clear yet.
In this paper, we report a stellar superflare detected with the Ground Wide Angle Camera (GWAC), along with rapid follow-ups with the F60A, Xinglong 2.16m and LCOGT telescopes. The effective temperature of the counterpart is estimated to be $2200\pm50$K by the BT-Settl model, corresponding to a spectral type of L0.
The $R-$band light curve can be modeled as a sum of three exponential decay components, where the impulsive component contributes a fraction of 23\% of the total energy, 
while the gradual and the shallower decay phases emit 42\% and 35\%  of the total energy, respectively.
The strong and variable Balmer narrow emission lines indicate the large amplitude flare is resulted from magnetic activity.
The bolometric energy released is about $6.4\times10^{33}$ ergs, equivalent to an energy release in a duration of 143.7 hours at its quiescent level. 
The amplitude of $\Delta R=-8.6 $mag ( or $\Delta V=-11.2$ mag),  placing it one of the highest amplitudes of any ultra-cool star recorded with excellent temporal resolution.  
We argue that a stellar flare with 
such rapidly decaying and huge amplitude at distances greater than 1 kpc may be false positive in searching for counterparts of catastrophic events such as gravitational wave events or gamma-ray bursts, which are valuable in time-domain astronomy and should be given more attention. 
\end{abstract}




\section{Introduction} \label{sec:intro}

Brown dwarfs (BDs) are  substellar objects   whose mass is between the mass of giant planets (11-16$M_\mathrm{J}$)\citep{2011ApJ...727...57S} 
and the stead hydrogen burning minimum mass (SHBMM, $\sim0.07M_\odot$, depending on the models and 
metallicity) \citep{2001RvMP...73..719B,2015A&A...577A..42B,2017MNRAS.468..261Z}. 
In steady of the energy released in fusion, pressure due to degeneracy of electrons is believed to resist gravitational collapse in the vast majority of BDs with $M<0.9$SHBMM. 
Although their formation mechanism is still in debate \citep{2014prpl.conf..619C}.
The popular star formation theories predict a great number of BDs in the universe  \citep{2001ApJ...556..830B,2002ApJ...567..304C,2003PASP..115..763C}. 
However, due to their small size, low temperature, and small luminosity,
only more than one thousand BDs have been identified from optical and infrared surveys
since the first discovery in 1995 
\citep{1998ASPC..134..405K,2023A&A...669A.139S,2011A&A...534L...7A,2018A&A...619L...8R}.

It is widely believed that the interiors of ultra-cool dwarfs (UCDs; spectral type $\geq$M7) 
is fully convective and is important laboratory for the study of stellar magnetic dynamics \citep{2012ARA&A..50...65L,2001ApJ...559..353M}. Strong kiloGuass magnetic field has been 
detected in a number of late M dwarfs through spectroscopy
\citep{2017NatAs...1E.184S,2010MNRAS.407.2269M,2007ApJ...656.1121R}
and in a batch of BDs through strong aurorae emission in GHz caused by the electron cyclotron maser instability
\citep{2022ApJ...932...21K,2018ApJS..237...25K,2016ApJ...821L..21R,2017ApJ...846...75P,2020ApJ...903...74R, 2008ApJ...684..644H}.

Although the periodically flares result from UCD's aurorae emission has been 
frequently detected in 
radio (see citations listed above), the flare rate in white-light is much lower than
that of early M-dwarfs by factors of 10-100
\citep{2018ApJ...858...55P,2020MNRAS.494.5751P,2013ApJ...779..172G,2017ApJ...845...33G,2016ApJ...828L..22S,2017ApJ...838...22G,2018RNAAS...2....8R,2019MNRAS.485L.136J,2021ApJ...922...78X}. Similar as occurred in the Sun and early M-dwarfs \citep{2017ApJ...851...91N}, 
these flares are believed to be 
result from the magnetic reconnection, which produce large amounts of energy in tens of seconds to hours. The super flares ($>10^{32}$ ergs) are intend to be accompanied with intense ultraviolet radiation and large coronal mass ejections at high velocities \citep{2012ApJ...760....9A,2021A&A...646A..34K,2021ApJ...916...92W}, which can have heavily impacts on the surrounding planetary atmosphere and the possible surface organisms \citep{2022MNRAS.509.5858H,2018SciA....4.3302R,2019AsBio..19...64T}.

In this paper, we report a super flare with an amplitude of $\Delta R=8.6$ mag 
on a L0 star detected in white-light by the GWAC system. Multi-wavelength
photometries and optical spectra during the gradual decay phase were
obtained for the flare. The total energy released in the bolometric energy is estimated to be about
$6.4\times10^{33}$ erg. This huge energy release makes the event
one of the strongest BD flares observed to date.

The ground based wide angle cameras (GWAC) is one of the main ground-based telescopes of the Chinese-French space-based SVOM mission \citep{2016arXiv161006892W}.
The main science of GWAC is to monitor a very wide sky (aiming with a total field-of-view (FoV) of $\sim 5000$ $\mathrm{deg}^2$ ) in a high cadence (the exposure time and readout time are 10 and 5 sec, respectively) with a shallow detection ability ($\sim 16$ mag in $R-$band). It focuses on the discovery of the optical transients associated to gamma-ray bursts \citep{2020GCN.29201....1X}, gravitational waves \citep{2020RAA....20...13T}, fast radio bursts \citep{2021ApJ...922...78X}, and stellar flares \citep{2021ApJ...909..106X,2021ApJ...916...92W,2022ApJ...934...98W}, and to explosive the short-duration transients in time-domain astronomy \citep{2019mmag.conf..260X}. 
GWAC will be composed of ten mounts in total. Each mount has four $JFoV$ cameras. Each $JFoV$ camera
has an effective aperture of 18 cm and a $f-ratio$ of $f/1.2$. Each camera is equipped with a $4K\times4K$ E2V back-illuminated CCD chip, giving a pixel scale of 11.7 arcsec. No standard filters are equiped. The wavelength coverage is from 0.5 to 0.85$\mu m$. Currently, 
four units with a total FoV of $\sim2200 \mathrm{deg}^2$ have been deployed in Xinglong observatory, Chinese Academy of Sciences, China.
 In the survey, each mount of GWAC monitors a designed sky 
region for at least 30 minutes \citep{2021PASP..133f5001H}.  The observations for each sky could be longer depending on the observation weather and the availability  of the instruments for each night. 

The paper is organized as follows. The discovery of the superflare is described in Section.\ref{sec:discovery}. Section.\ref{sec:follow-ups} reports the rapid follow-ups
by both photometry and spectroscopy. The results are presented in Section.\ref{sec:res}. Section.\ref{sec:discussion} gives discussion and conclusion.

\section{Discovery of the super flare}\label{sec:discovery}

\begin{figure}
 \centering
  
   \subfigure[GWAC image before trigger time T0]{
   \label{fig:subfig:A}
   \includegraphics[width=0.2\textwidth,frame]{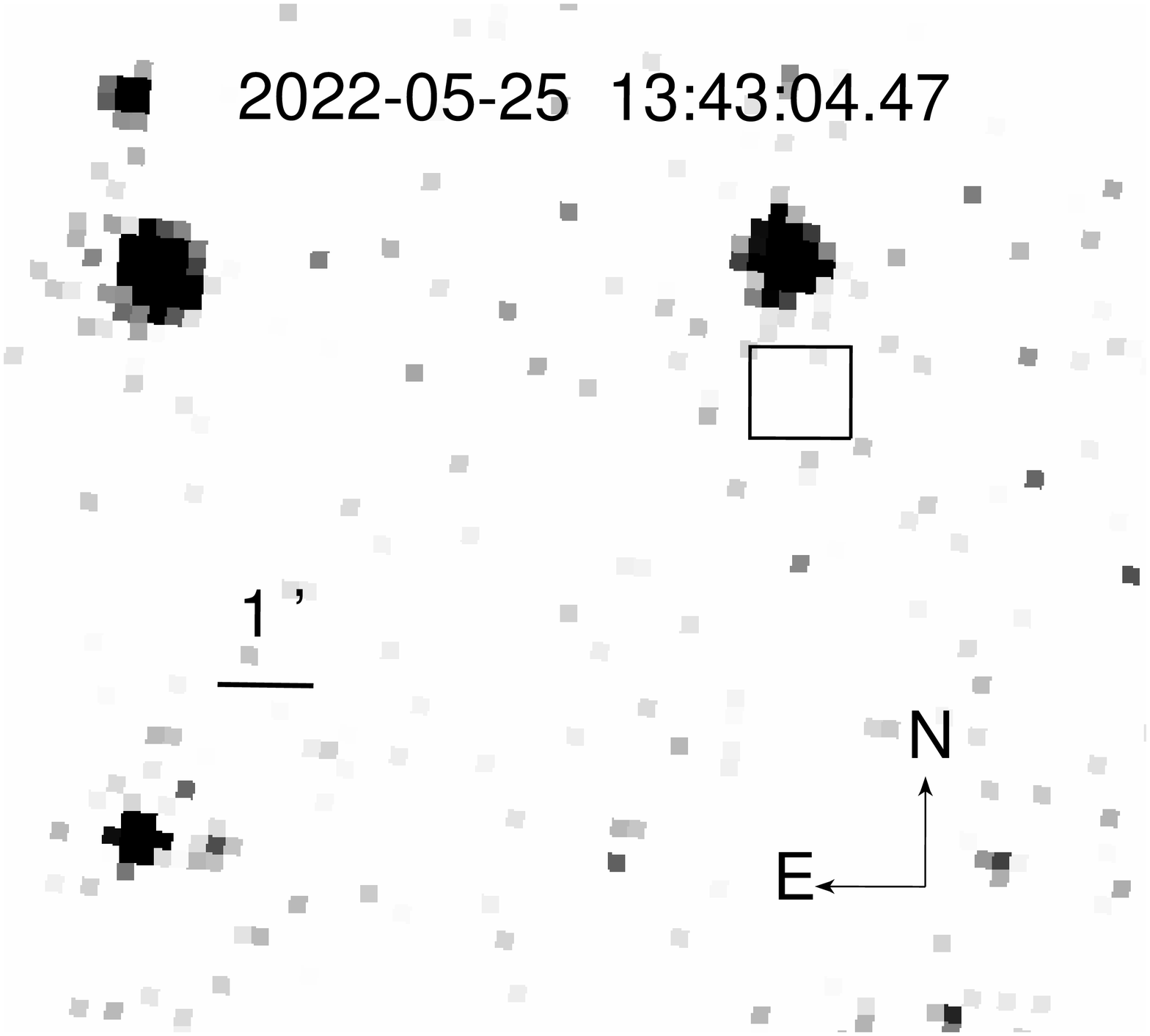}
   }
   \hspace{0.1in}
   \subfigure[GWAC image at trigger time T0 ]{
   \label{fig:subfig:B}
   \includegraphics[width=0.2\textwidth,frame]{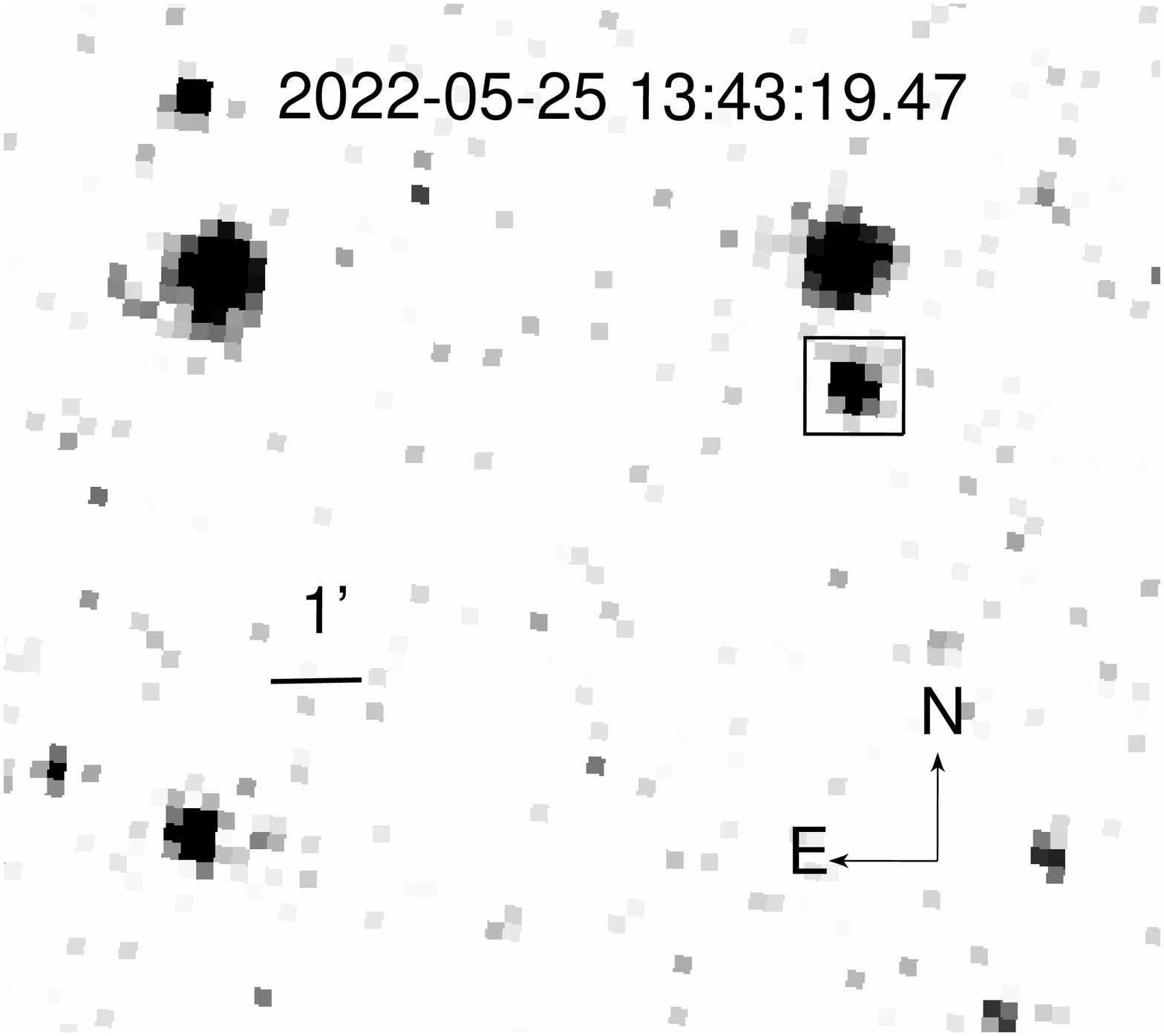}
   }
   
   \subfigure[F60A image]{
   \label{fig:subfig:C}
   \includegraphics[width=0.2\textwidth,frame]{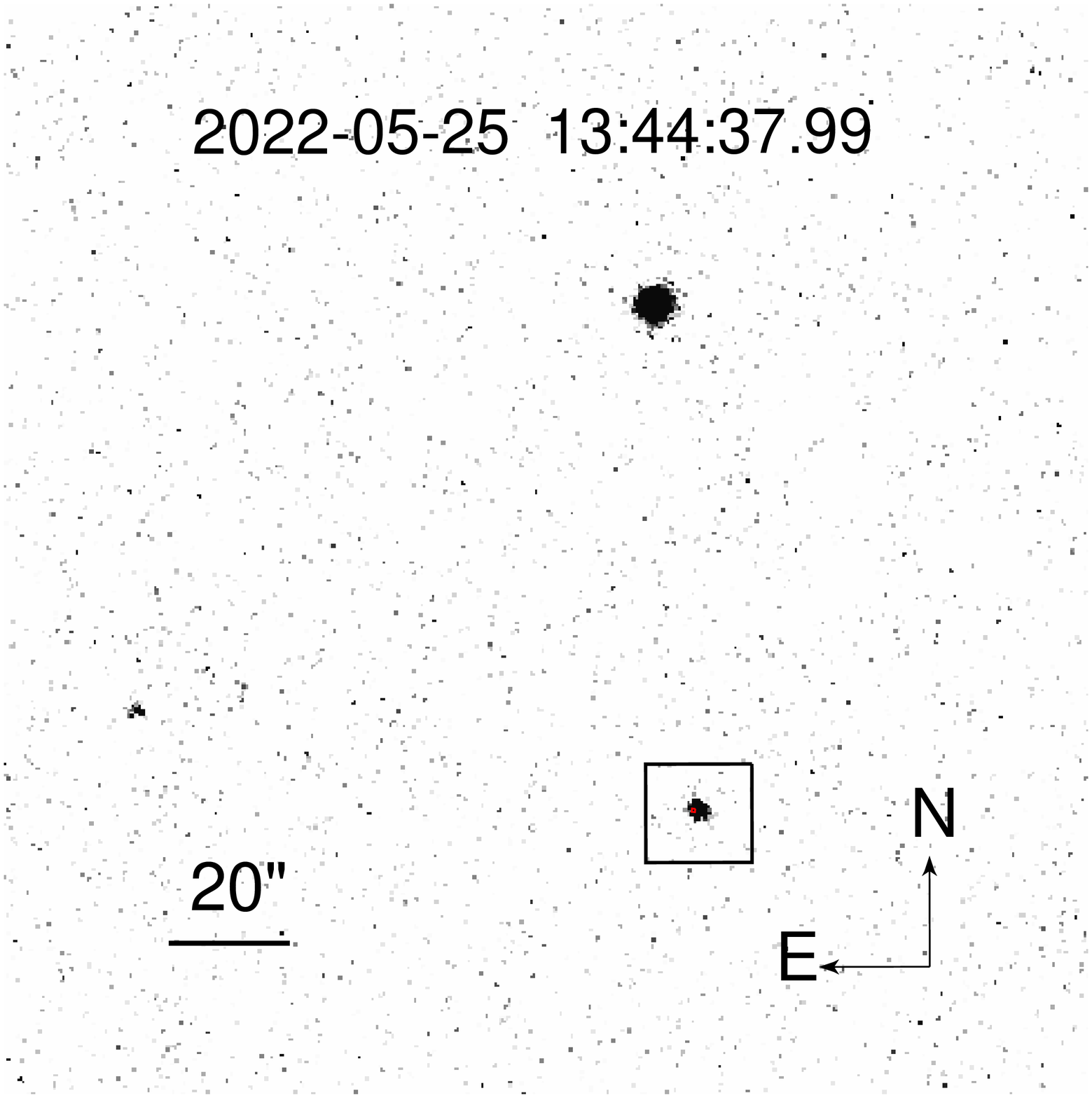}
   }
   \hspace{0.1in}
   \subfigure[SDSS image]{
   \label{fig:subfig:D}
   \includegraphics[width=0.2\textwidth,frame]{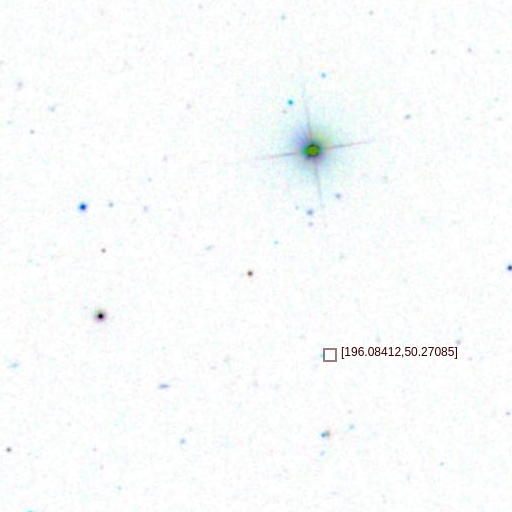}
   }
  
\caption{
Finding charts for GWAC220525A detected by GWAC,F60A and SDSS DR14.
The localization of the flare is marked with a black box in each panels.
\it Panel (a) and (b): \rm GWAC images labeled with the universal time, which were before and near the peak of the event, respectively. 
The upper is North and the left is EAST.
\it Panel (c): \rm One image obtained by F60A, 79 seconds after the trigger time. The universe time for the measurement is also labeled. 
\it Panel (d): \rm
reference image derived from the SDSS DR13 survey for a comparison. 
\label{fig:findingchart}}
 \end{figure}

On 2022 May 25 UT13:43:19, a fast fading transient was discovered by the GWAC online pipelines for a very bright optical emission (GWAC220525A)
during its monitoring of the sky from 12:46:21 to 17:40:49 UT.  
The coordinates of the source measured from the GWAC images are 
R.A.=13:04:20.148 DEC=+50:16:14.53, J2000. The corresponding astrometric precision is 
about 2.0 arcsec (1$\sigma$). 
There is no any counterpart in not onlyour reference image which was obtained 
at 2020, Jan. 2 UT18:31:57,  but also the successive frames taken by GWAC before the event.
The finding charts before and during the first detection from GWAC as well as the image from the follow-up telescope F60A are displayed in Figure.\ref{fig:findingchart}.  
The SDSS  image is also shown for the comparison.  

A standard aperture photometry was carried out at the 
location of the transient and for several nearby bright 
reference stars by using the IRAF\footnote{IRAF is distributed by the National Optical Astronomical Observatories, which are operated by the Association of Universities for Research in Astronomy, Inc., under cooperative agreement with the National Science Foundation.} APPHOT package.  
The corrections of bias, dark and flat-field were performed in a standard manner.
The finally calibrated brightness of the source was obtained by using the SDSS catalogues through the Lupton (2005) transformation\footnote{\url{http://www.sdss.org/dr6/algorithms/sdssUBVRITransform.html\#Lupton2005} ($R = r - 0.2936(r - i) - 0.1439$; $\sigma=0.0072$).}.

\section{Follow-ups in Photometry and Spectroscopy}\label{sec:follow-ups}
\subsection{NAOC-GuangXi F60A observations}
Once the flare was triggered by the GWAC real-time pipeline after passing all the filters, including the automatically check of the point spread function, the database of the known bright minor planets, the possibility of the moving among our images, it was immediately follow-uped by F60A via a dedicated real-time automatic transient validation system (RAVS)\citep{2020PASP..132e4502X}.  This system is established not only to confirm candidates, but also to obtain a light-curve with optimized adaptive sampling and exposure 
based upon the brightness estimated from the evolution trend. 
For the case of GWAC220525A, rapid follow-up by F60A started at 1.3 min after the first detection of GWAC, and lasted for about 5 hours with an exposure time for each frame in the range  from 30 to 200 seconds. 
More accuracy of the localization of the source was derived with 
R.A.=13:04:20.2, DEC=+50:16:14.5, J2000, with a precision of 0.1 arcsec (1$\sigma$),
thanks for the high spatial resolution of F60A relative to GWAC images. 

All the data reduction are performed in the standard manner including the correction of bias, dark and flat-field. The photometry calibration is carried out against the SDSS catalogs. 

\subsection{Xinglong 2.16m telescope observations}
Xinglong 2.16m optical telescope \citep{2016PASP..128k5005F} was triggered via a Target of Opportunity request  at UT14:24:32, 41 minutes after the event. Two spectra were obtained by using the Beijing Faint Object Spectrographa and Camera (BFOSC)\footnote{The BFOSC camera is equipped with a back-illuminated E2V55-30
AIMO CCD.}. The exposure time are 10 min and 20 min. With a slit width of 1.8 arcseconds oriented in the south–north direction, the corresponding spectral resolution is $\sim$10\AA, when grating
G4 was used. This setup results in a wavelength coverage of
3850\AA–8000\AA. The wavelength calibration was performed with the
iron–argon comparison lamps. 
After a bias subtraction and flat-field correction,
standard procedures were
adopted to reduce the two-dimensional spectra by using the IRAF package, 
The extracted one-dimensional spectrum was then
calibrated in wavelength and in flux by the corresponding
comparison lamp and standard calibration stars, respectively.

After the spectroscopic observations, additional photometric monitors, 
lasting for about 4 hours in total, were carried out by the BFOSC. 
The exposure time was set from 100 to 400 seconds according to the ability of the 
telescope\footnote{The ability considered here includes not only the diameter and  efficiency, but also the tacking stability of the telescope as well as the observation conditions that night.} and the fading brightness estimated by the RAVS. 
All the data are analyzed in the standard manner including the correction of bias and flat-field. Image stacking was then performed to increase the signal-to-noise ratio for 
subsequent photometry, which results in 25 measurements in the $R-$band.
The photometry calibration is carried out against the SDSS catalogs in the same field again. \rm

\subsection{Photometry by LCOGT}
The follow-up photometries were performed by Las Cumbres Observatory Global Telescope (LCOGT) \citep{2013PASP..125.1031B} with three runs from 
May 25 to 27. 
20 $R$-band frames were obtained in total.  The exposure time of each image 
was set to be 300 seconds, which is determined from the brightness estimated from 
the trend of the brightness estimated from the measurements by both F60A and Xinglong 2.16m telescopes, and the limiting magnitude of the LCOGT telescopes. 
The correction of  bias, dark and flat-field was carried-out with the IRAF software. The astrometric calibration was performed with the Astronomey.net \citep{2010AJ....139.1782L}. The flux calibration was also performed with the same manner adopted for the Xinglong 2.16m and F60A observations. 
The source is clear detected by stacking the images on each night. 

\begin{figure}
 \centering
 \includegraphics[width=0.5\textwidth]{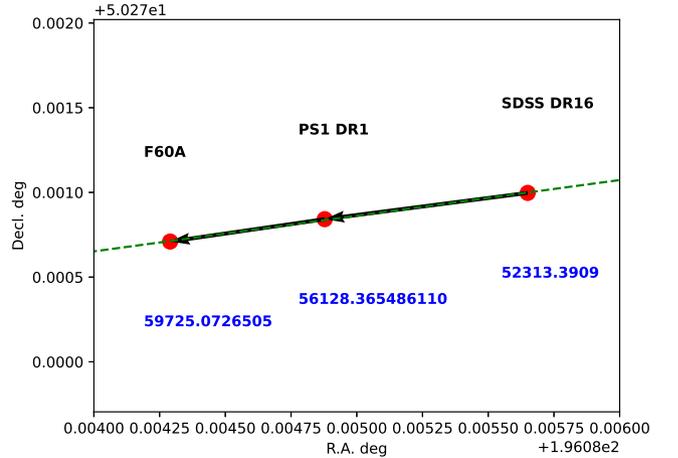}
\caption{
 The celestial coordinates derived from SDSS DR16, Panstarrs DR1 and F60A (this work) are plotted by the red circles. The green dashed line and thick black arrows mark the locus on the sky by a linear fitting. The corresponding modified Julian days are labeled in blue.    
The values of the proper motion are deduced to be $\mu_{\text{RA}}=-241$ and $\mu_{\text{Dec}}=-50\ \mathrm{mas\ yr^{-1}}$, with an maximum of  uncertainties of 10\%.   
\label{fig:pos}}
 \end{figure}

 \begin{figure}
 \centering
  \includegraphics[width=0.5\textwidth]{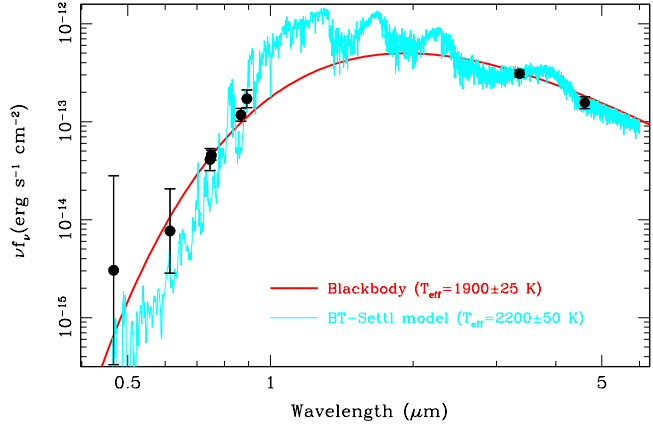}
\caption{
Spectral energy distribution for the source at the quiescent state. The data shown by the
black circles are derived from the public archive (WISE, PanStarrs and SDSS surveys, Table 1 in this work). 
The red and cyan curves show the best-fit blackbody spectrum with a 
temperature of $1900\pm50$K and BT-Setti model atmosphere with a temperature
of $2200\pm50$K, repsectively 
\label{fig:sed}}
 \end{figure}

\section{Result} \label{sec:res}
\subsection{The quiescent counterpart and the nature of the event}
With the astrometry, we cross-match the source with the existing survey catalogs
through the VizieR
Service\footnote{\url{https://vizier.u-strasbg.fr/viz-bin/VizieR}}.
Only one source, hereafter named SDSS\,J1304 for short, can be founded in 
SDSS \citep{2000AJ....120.1579Y}, Wide field Infrared Survey Explorer (WISE) \citep{2010AJ....140.1868W}, PanSTARRS DR1 catalog (PS1) \citep{2016arXiv161205560C} within a search radius of 4 arcseconds.
No any counterparts can be found at other catalog including Gaia catalog \citep{2018A&A...616A...1G}.
The queried key parameters are presented in Table.\ref{Survey}.
We plot the celestial coordinates measured by F60A, SDSS and PanStarrs together in Figure.\ref{fig:pos}, which shows an evident proper motion
in the past $\sim$20 years. 
The values of the proper motions are deduced to be $\mu_{\text{RA}}=-241$ and 
$\mu_{\text{Dec}}=-50\ \mathrm{mas\ yr^{-1}}$. 
The uncertainties of the proper motions estimated above shall be less than 10\% considering the uncertainties of localization and the observation time for the measurements. 

All magnitudes in Table.\ref{Survey} are converted to flux density, and the spectral energy distribution (SED) in the quiescent state is built and displayed in  Figure.\ref{fig:sed}. 
The SED is modeled with a black body (BB) spectrum and the BT-Settl model atmospheres available for stars, BDs and planets \citep{2014IAUS..299..271A}  by a $\chi^2$ minimization. The reduced $\chi^2$ is 0.90 and 1.18 for the blackbody and BT-Settl model, 
respectively. Both best-fit models are overplotted in Figure \ref{fig:sed}.  The effective temperature $T_{\text{eff}}$ is obtained to be $1900\pm25$K for the 
blackbody model and $2200\pm50$K for the BT-Settl model. 
The latter temperature is adopted in this study, which corresponds to a spectral type 
of M9-L1 according to the relationship between $T_{\text{eff}}$ and the spectral type for low-mass stars \citep{2015ApJ...810..158F}. We take L0 as its spectral type for the following analysis.

Due to its faintness, there is no any reported 
parallax about the distance of SDSS\,J1304. 
Based on the relation between absolution magnitude $M_{\text{W2}}$ and spectral type
\citep{2015ApJ...810..158F},  $M_{\text{W2}}\sim10.0$ mag is obtained. 
With the distance modulus of 4.622, the distance of the source from us could be deduced 
to be $d=84\pm5$ pc  after taking the uncertainties of the relation into account.

Taking the first detection by the GWAC is at the peak of the flare (see Section 4.2), the flare amplitudes
are calculated to be $\Delta V=-11.2$ mag and $\Delta R=-8.6$ mag by assuming a 
balckbody emission with a temperature of $\sim10^{4}$K \citep{2013ApJS..207...15K,2022arXiv220411059M} at the peak. 
With the transparent curves and the zero points of individual filter,
the quiescent emission level in $V-$ and $R-$bands used in the calculation 
is estimated to be $V_{\mathrm{q}}=25.5\pm0.1$ mag and $R_\mathrm{q}=22.9\pm0.1$ mag,
respectively, by an integration of the best-fit BT-Settl model shown in Figure.\ref{fig:sed}.

\begin{figure}
 \centering
 \includegraphics[width=0.5\textwidth]{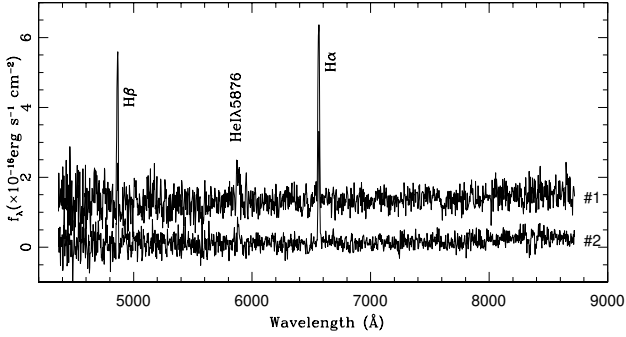}
\caption{
The two spectra obtained by the  2.16 m telescope at Xinglong
Observatory, China, during the gradual decay phase. The observations started at 41 min after the event.  \#1 is the first spectrum with an exposure time of 600 sec. and \#2 show the second spectrum with an exposure time of 20 min. Emission line features are labeled in the figure, which confirms a high chromospheric activity of the source.
\label{fig:spec}}
 \end{figure}

The two spectra obtained by the 2.16m telescope at the end of the gradual decay phase 
(see Figure.\ref{fig:lightcurve} for the details) are 
displayed in Figure.\ref{fig:spec}.
Both spectra show emission lines of 
$\mathrm{H\alpha}$, $\mathrm{H\beta}$, and \ion{He}{1}$\lambda5876$,  
which are commonly detected during a dMe's flare \citep{2013ApJS..207...15K} and considered to be associated with chromospheric activity. This result suggests that this kind of activities also take place in the BD studied here.  We measure 
the flux of $\mathrm{H\alpha}$ and $\mathrm{H\beta}$ emission lines 
by a direct integration within the wavelength region 6544-6576\AA\ and 
4850-4874\AA, respectively. The measured fluxes, along with the uncertainties\footnote{The statistic error of line $\sigma_l$ is determined by the method  given in Perez-Montero \& Diaz (2013): $\sigma_l = \sigma_c\sqrt{N[1+EW/(N\Delta)]}$ , where $\sigma_c$ is the standard deviation of continuum in a box near the line, $N$ the number of pixels used to measure the line flux, $EW$ the equivalent width of the line, and $\Delta$ the wavelength dispersion in unit of \AA\ $\mathrm{pixel^{-1}}$.}, are listed in Table.\ref{spec_line}\footnote{It is noted that the line fluxes reported here include the quiescent level, since the quiescent brightness of the source is too faint, and the corresponding SDSS spectrum at the quiescent state is too noise to extract any information of the source}. 
Both spectra return a Balmer decrement of $\mathrm{H\alpha/H\beta}\approx1.3$. 
This decrement can be understood by an emission from an optically thick plasma with 
a temperature around 5000K according to the local thermal equilibrium calculation \citep{2012A&A...537A..94S}.

\begin{table}
\begin{center}
\caption{Properties of SDSS\,J1304 (the quiescent counterpart of GWAC\,220525A) extracted from various surveys. "Offset" means the angular distance from the position given by the survey to the one measured by the F60A.}
\begin{tabular}{lll}
\hline Parameter & unit & Value \\
(1) & (2) & (3) \\
\hline  \multicolumn{3}{c}{SDSS\,J130420.55+501615.5}\\
\multicolumn{3}{c}{\citep{2020ApJS..249....3A}} \\
\hline R.A. & degree &  196.085650 \\
Decl. & degree & +50.270997d \\
$u$ & mag & $24.847 \pm 0.652$ \\
$g$ & mag & $25.023 \pm 0.473$ \\
$r$ & mag & $23.714 \pm 0.344$ \\
$i$ & mag & $21.083 \pm 0.064$ \\
$z$ & mag & $19.229 \pm 	0.045$ \\
JD & day & 2452313.3909 \\
Offset  & arcsec & 1.434 \\
\hline \multicolumn{3}{c}{Pan-Starrs DR1 (168321960847785676)}\\
\multicolumn{3}{c}{\citep{2016arXiv161205560C}}  \\
\hline R.A. & degree & 196.084877950  \\
Decl. & degree & +50.270842040d \\
$i$ & mag & $21.1139 \pm 0.0266$ \\
$z$ & mag & $19.5545 \pm 0.0327$ \\
$y$ & mag & $18.6101 \pm	0.0274$ \\
JD & day & 2456128.365486110 \\
Offset & arcsec & 3.295 \\
\hline \multicolumn{3}{c}{WISE\,J130420.42+501615.1} \\
\multicolumn{3}{c}{\citep{2013wise.rept....1C}} \\
\hline
R.A. & degree & 196.085124  \\
Decl. & degree & +50.270881 \\
$W 1$ & mag & $14.869 \pm 0.031$ \\
$W 2$ & mag & $14.622 \pm 0.05$ \\
Offset  & arcsec & 1.870 \\
\hline
\label{Survey}
\end{tabular}
\end{center}
\end{table}

\begin{table}
\begin{center}
\caption{Emission line measurements of the spectra of GWAC\,220525A displayed in the Figure.\ref{fig:spec}.
The emission flux is in the unit of ($\mathrm{10^{-15}\,erg\,s^{-1}\,cm^{-2})}$}
\begin{tabular}{ccc} 
\hline 
Line  &   Flux(\#1)  &  Flux(\#2) \\
(1) & (2) & (3)\\
\hline 
$\mathrm{H\alpha}$ &       8.1$\pm$0.2   & 4.5$\pm$0.1 \\ 
$\mathrm{H\beta}$      &   6.1$\pm$0.4 & 3.4$\pm$0.6 \\
\hline
\label{spec_line}
\end{tabular}
\end{center}
\end{table}

\begin{table}
\begin{center}
\caption{Fitted parameters of GWAC\,220525A used in Equation.\ref{func2}.
}
\begin{tabular}{cc} 
\hline 
$k_1$ & 0.715$\pm$0.001 \\
$\alpha_1$& 0.767$\pm$0.001 \\
$k_2$ & 0.255$\pm$0.0002 \\ 
$\alpha_2$& 0.148$\pm$0.0002 \\
$k_3$ & 0.0409$\pm$0.00003 \\
$\alpha_3$ & 0.0251$\pm$0.00007 \\

\hline
\label{fitting_parameters}
\end{tabular}
\end{center}
\end{table}

\subsection{Optical light curve}

\begin{figure}
 \centering
 \includegraphics[width=0.5\textwidth]{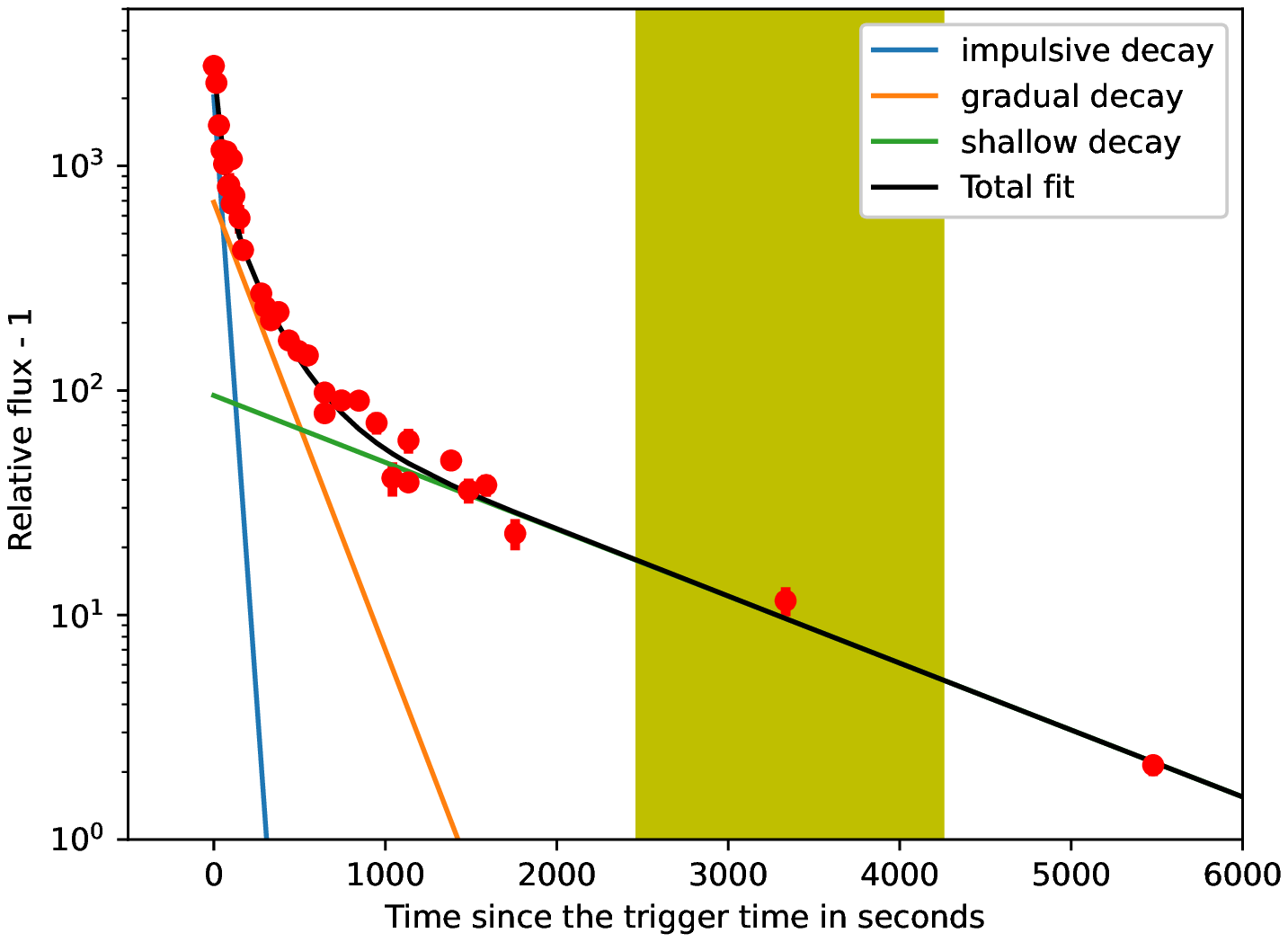}
 \includegraphics[width=0.5\textwidth]{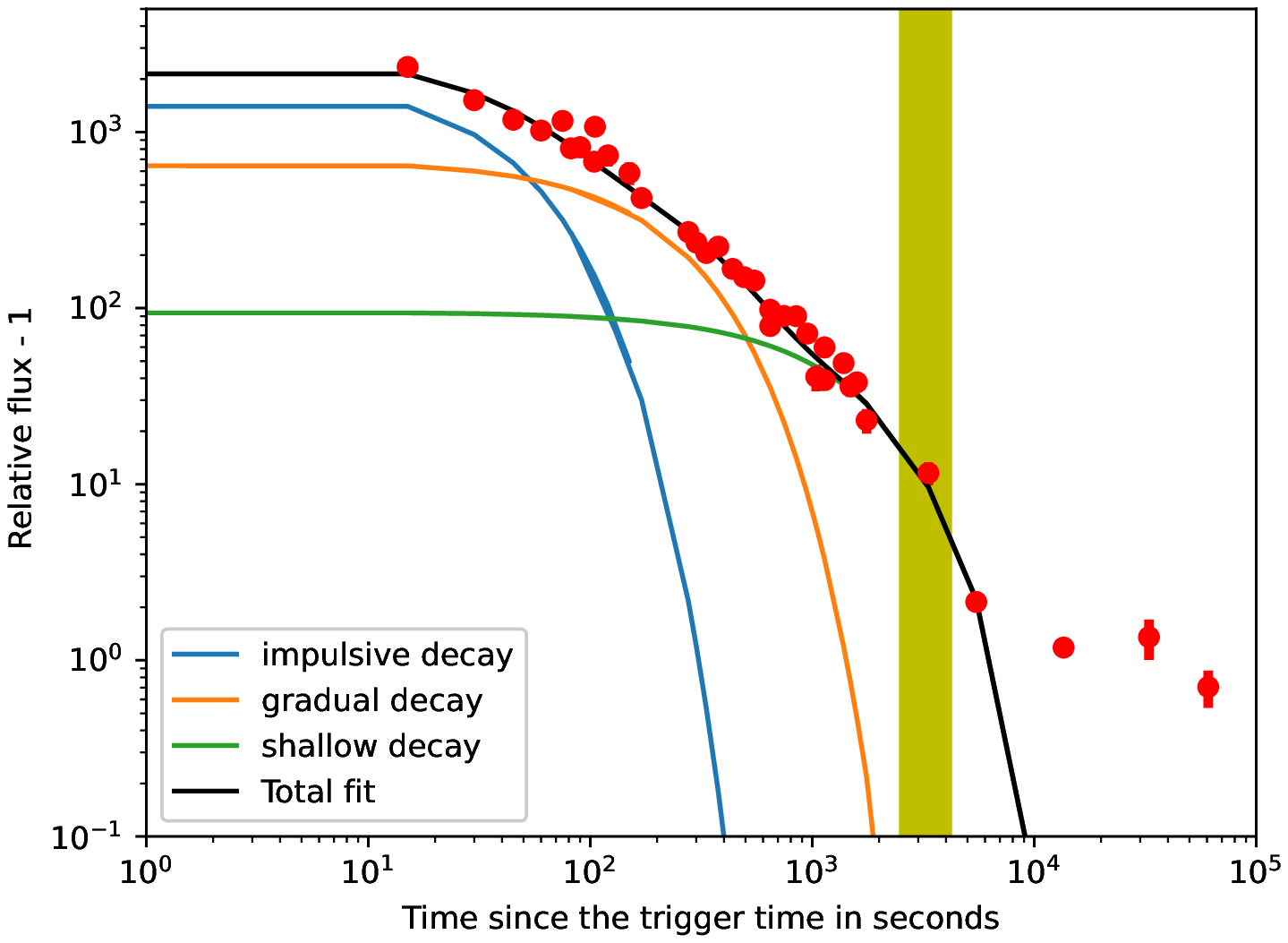}
\caption{
The $R$-band light curve of GWAC\,220525A observed by  GWAC, F60A, Xinglong 2.16m, and LCOGT telescopes and its modeling. X-axis is the seconds since the trigger time given by the GWAC. Y-aix is the flaring flux normalized by
the quiescent flux level. 
The last four points are quite close to the quiescent flux level, which are not included in the light-curve fitting after taking their uncertainties into account.  The best-fit model and the three exponential components 
are overplotted by the solid lines in different colors. The vertical shadowed region from bottom to top marks the epoch when the two spectra were 
obtained by the Xinglong 2.16m telescope. 
The upper panel is the zoomed bottom panel without the display of the measurements after 6000 seconds for the clarity of the modeling during the early phase. 
The y-axis for the upper panel is in the linear scale while the y-axis for the bottom panel is in the Logarithmic scale.
\label{fig:lightcurve}}
 \end{figure}

The $R-$band light curve built from all the measurements from GWAC, F60A, 2.16m and LCOGT telescopes is ploted in Figure.\ref{fig:lightcurve}. 
 We argue that the first detection of the flare by GWAC is 
close to the peak of the light curve, both because of the non-detection in the frame taken before the the first detection and because 
the first detection of the flare is the brightest in the light curve.
The missing of the rising phase could be resulted from either a very fast energy release or the limited detection ability of the GWAC. 
Given the detection limit and survey cadence, the flare brighten
by 1.5 magnitude in 15 seconds near the peak in the $R-$band.

By following our previous studies
\citep{2021ApJ...909..106X,2021ApJ...916...92W,2022ApJ...934...98W}, a sum of a set of exponential functions 
is adopted to describe the decaying phase: \rm
\begin{equation} 
\frac{F_{\text {decay}}}{F_{\text{amp}}} = \sum_{i=1}^{n} k_i e^{-\alpha_i t }    
\label{func2}       
\end{equation}
where $F_{\text {decay}}$ is the flux during the decay phase, and 
$F_{\text {amp}}$ the amplitude flux of the flare which is determined at the peak time. $\alpha$ is the decaying slope.  $n$ stands for the number of the components which is needed to account for the behavior of the light curve during the decaying phase, and $k$ is the amplitude of each component. 
In GWAC\,220525A, our inspection by eyes shows that there are four measurements deviating the expected exponential decay at the end of the light curve (i.e., $\sim10^{4}$ seconds since the peak). \rm 
However, considering that the brightness derived from the four measurements is quite close to the 
quiescent level, we do not include them in our light-curve fitting and 
analysis. Briefly speaking, three exponential components are required 
to reproduce the observed light curve. 
In the light-curve fitting, $F_{\mathrm{amp}}$ is fixed to be 2788.97, and 
a bayesian information criterion (BIC) is adopted to test weather the fitting result by the model is over-fitted\footnote{BIC=1476.6 and 475.8,  284.9, 284.1 for model with one, two, three and four components respectively. More small value for BIC, more better for the fitting. On the other hand, given the BIC value for three and four components is comparable, three-component model is adopted in this work.} 
 The best-fit model associated with a resulted $\chi^2 = 95.2$ with a degree of freedom of 28 is overplotted in Figure 5. 
Table.\ref{fitting_parameters} summarizes the fitted parameters of the three components.

\subsection{Energy budget}
Following the approach adopted in previous studies \citep{2013ApJS..207...15K,2021ApJ...909..106X}, the total energy $E_R$ in the $R$-band can be estimated by the formula $E_{R}=4\pi r^2\times F_{R,\mathrm{q}}\times ED$, where
the quiescent flux $F_{R,\mathrm{q}} = 2.4\times10^{-15}$  erg  cm$^{-2}$ s$^{-1}$,
and the distance is $r=84$ pc.
The equivalent duration $ED$ of a flare is defined to be the time needed to emit all the flare energy at a quiescent flux level \citep{2013ApJS..207...15K}. 
By integrating the modeled light curve over the range  from the start to the end of the flare, 
$ED$ is estimated to be $\sim 5.17\times10^{5}$ seconds, 
or 143.6829 hours. 
The energy $E_{R}$ is therefore evaluated to be $(1.07\pm0.1)\times10^{33}$ ergs. Strictly speaking,
the method adopted here to estimate energy is only valid for the case in which the flare spectrum is same to the one 
in the quiescent state. Although this requirement is not fully consistent with fact \citep{2021ApJ...909..106X}, the resulted systematic on the energy due is only $\sim8\%$ according to our experience. 
The bolometric energy is $(6.4\pm0.7)\times10^{33}$ ergs by assuming
the temperature of the flare at the peak time is $10^4$K \citep{2013ApJS..207...15K}.
The fractions of the emitted energy are: 
23\%,  42\% and 35\% in the impulsive-, gradual-, and shallow-decay phases, respectively.

\subsection{Constraints on magnetic field}
 
The maximum of the vertical magnetic field $B_z^{\mathrm{max}}$ can be estimated by the relations given in \citep{2013A&A...549A..66A}
\begin{equation}
   E_{\mathrm{bol}}=0.5\times10^{32}\bigg(\frac{B_z^{\mathrm{max}}}{1000\mathrm{G}}\bigg)^2\bigg(\frac{L^{\mathrm{bipole}}}{50\mathrm{Mm}}\bigg)^2\ \mathrm{erg}
\end{equation}
where $E_{\mathrm{bol}}$ and $L^{\mathrm{bipole}}$ is the bolometric flare energy and the linear separation between bipoles, respectively. 
The energy released for GWAC220525A returns a magnetic field about 1.3 kG, 
when the $L^{\mathrm{bipole}}$ is taken as the maximum distance $\pi R$, where the radius $R$=0.1\(R_\odot\) is taken  as a typical radius of a L0 type star.  A kiloguass magnetic field has, in fact, been detected frequently in M to T dwarfs\citep{2008ApJ...684..644H,2010MNRAS.407.2269M,2016ApJ...821L..21R,2018ApJS..237...25K,2017NatAs...1E.184S,2007ApJ...656.1121R,2017ApJ...847...61B}.
In addition, a strong field
of 3.6–4.7 kG has been inferred in the white-light superflare of a M9-type star detected 
by the GWAC (i.e., GWAC\,181229A\citep{2021ApJ...909..106X}).

\section{Discussion and conclusion} \label{sec:discussion}

We here present a powerful white-light flare GWAC\,220525A from a L0 star triggerred by the GWAC intense observations at a 15-second short cadence.  
This is the second white-light stellar flare of very high amplitude ($\Delta R = -8.6$ mag or $\Delta V = -11.2$ mag) and large energy release of an ultra-cool star independently detected by GWAC since 2018.  
The amplitude of GWAC\,220525A  is comparable with those previously reported in 
ultra-cool stars, such as the
GWAC\,181229A from a M9 star with $\Delta R = -9.5$ mag \citep{2021ApJ...909..106X},
the powerful stellar flares from an early L dwarf ASASSN-16ae \citep{2016ApJ...828L..22S} with $\Delta V < -11$ mag,  
L1 dwarf J005406.55-003101.8 with $\Delta V\sim -8$ mag \citep{2017ApJ...838...22G}, ASASSN-18di with $\Delta V = -9.8$ mag \citep{2018RNAAS...2....8R}，L2 dwarf 
ULAS\,J224940.13-011236.9 with $\Delta V \sim -10$ mag \citep{2019MNRAS.485L.136J}, and a superflare with the largest amplitude of $\sim 6.3$ mag detected from \emph{Kepler/K2} mission on a L5 dwarf \citep{2020MNRAS.494.5751P} .  

\subsection{Detection probability of the rare events}
The cumulative flare frequency $\nu$ is related with energy $E$ as $\log{\nu}=\alpha+\beta\times\log(E_{32})$ \citep{1976ApJS...30...85L} , where $\nu$ and 
$E_{32}$ are in unit of $\mathrm{day^{-1}}$ and $1\times 10^{32}$ erg, respectively. $\alpha$ and $\beta$ are the normalization and slope, respectively.
There is evidence that the values of $\beta$ 
are similar with each others between BDs, the Sun, and other stars, but the normalization parameter is different \citep{2013ApJ...779..172G,2017ApJ...838...22G}.
However, other studies presented that the slopes are much shallower in cooler stars
\citep{2018ApJ...858...55P,2020MNRAS.494.5751P}. 
 Based on $\alpha=-1.35\pm0.06$ and $\beta = -0.59\pm 0.09$, which is derived from 
the flares in the energy range between $10^{31}$ and $2\times 10^{32}$ erg in L1 dwarf W1906+40 \citep{2013ApJ...779..172G},
the rate for GWAC\,220525A is predicted to be about 1.45 per year for a flare with
energy larger than $6\times10^{33}$ erg. 
One has to be caution that this result is sensitive to the parameters we used, which might be not the fact for the source studied in this work.

The detection limit of the GWAC is typical of 15.0 mag in the $R-$band, which means that there is a gap from the detection ability to the maximum brightness of GWAC\,220525A for about 1.6 magnitudes. This difference in magnitude corresponds to an energy 
difference of 4.3 times, which means that only the flares from this dwarf with an energy larger than $\sim1.4\times10^{33}$ erg could be detected by the GWAC single images.  
It therefore results in a detection rate of about 3.44 per year,
which is comparable with the estimates of the occurrence rates on the L dwarfs\citep{2020MNRAS.494.5751P}, i.e., evergy 2.4 yr for a superflare with energy of $10^{33}$ erg, and every 7.9 yr for a superflare of energy of $10^{34}$ erg. 
Actually, we have carried out an off-line search of any other flares from this source in the GWAC achieved data obtained from Jan. 2018 to June 2022. In total, we searched 92,161 images. The  exposure time for each frame has been fixed to be 10 seconds in our survey, corresponding to a total effective observation time of 256 hours for this field. Only the flare GWAC\,220525A was detected in our off-line search. Regardless of the variation of the detection ability during these observations, the detection possibilities of such rare event from this source shall be less than 0.0039 per hour. This non-detection of other superflares supports the low flare rate estimated from the L1 dwarf W1906+40 \citep{2013ApJ...779..172G}. More monitor needs to be performed to confirm the above estimation.

\subsection{Complex of light curve and the configuration of magnetic fields}
The light curve of the superflare in GWAC\,220525A shows several features which make its decay deviates from an exponential shape. Several components are needed to model the whole shape of the light curve. 
The continuous cooling of the flaring plasma is interrupted obviously in this event, suggesting that the whole flare originates from multiple active regions. These results suggest that a complex structure of the flare with
several loops superimposed consecutively, rather than a single loop\citep{2011SSRv..159...19F}.

The multiple phases in the light curve could reflect an evolution of the configuration of the magnetic fields in the corona. 
During the impulsive phase of GWAC\,220525A, the emitted energy is only 23\% of the total energy. 
The dominated energy is released in the later phases.
 For the gradual phase in this work, the energy faction is up to  42\%,  indicating new powerful magnetic reconnection occures continuously as the flare
arcade caused by  the chromospheric evaporation  effect in the hot plasma in the framework of the 
“CSHKP model” \citep{1964NASSP..50..451C,1966Natur.211..695S,1974SoPh...34..323H, 1976SoPh...50...85K}. 
In the current event, an additional decay phase with a much shallower decay index and 
an energy faction of about 35\% is detected after the gradual phase. 
This kind of component is also detected in other superflare GWAC\,181229A \citep{2021ApJ...909..106X}.  This additional energy could be generated from an 
ongoing slow magnetic reconnection and its associated heating \citep{1979SoPh...61...69M,1989SoPh..120..285F}, either above the flare
arcade, or conceivably between individual tangled strands of the arcade as they
shrink down \citep{2011SSRv..159...19F}.

\subsection{The possible accompanying CME}
Many observations present that more powerful a solar flare, 
more intense the associated coronal mass ejections (CME) will be \citep{2008ApJ...673.1174Y,2011SoPh..268..195A,2012LRSP....9....3W,2012ApJ...760....9A}. This linkage might be valid for the stellar flares \citep{2019ApJ...877..105M}. 
A few CME candidates accompanying by the stellar flares have been reported in 
previous studies \citep{1990A&A...238..249H,2021A&A...646A..34K,2021ApJ...916...92W,2022ApJ...933...92C}.  
A superflare such as reported in this work is expected to be associated with a powerful CME, though there are some debates on the possible suppression on the CME in late M or cooler stars \citep{2018ApJ...862...93A,2019ApJ...874...21F}. Considering the habitable zone of a L0 BD is located at $\sim$0.07 AU from the host \citep{2016PhR...663....1S}, 
the possible CME, UV emission and high energy particles associated with the superflare would have very heavy impacts not only on the fully co-evolved atmosphere but also on the life \citep{2017ApJ...846...31C,2018AsBio..18..663S} on the planets if existing around this dwarf.

Unfortunately, the grism G8  with a spectral resolution of
$\sim 2$\AA\ 
was not available for the Xinglong 2.16m telescope on the night. The low spectral resolution ($\sim 10$\AA) of the grism G4 we used does not allow us to detect a possible CME as performed in our previous observations \citep{2022ApJ...934...98W}, unless the velocity of the outflow along the 
line-of-sight is higher than 1300 $\mathrm{km\ s^{-1}}$.

\subsection{Possible "Frog" in  time-domain astronomy}

The era of time-domain astronomy has been coming. Various facilities with wide FoV in the word-wide are built to discover the counterparts associated to the gravitational wave, gamma-ray burst, fast radio burst, shock break-out emission from supernove, and other cosmic violent explosions. 
Among these activities,
it is no doubt a tough task to identify an optical counterpart of an event with a poor localization \citep{2020LRR....23....3A}, such as a gravitational wave event.  A localization with an uncertainty of hundreds to thousands degrees \citep{2017PhRvL.119p1101A} makes it not realtise to have intense and deep search for most of facilities in the worldwide if more sky-maps are intent to be covered\citep{2020RAA....20...13T,2022ApJ...926..152F}. A high-amplitude stellar flare  with a rapid decay slope from a very faint dwarf such as GWAC\,220525A could make it a false-positive for these cosmic event. For example, the counterpart of this flare is estimated at a distance of 84 pc. If it was more distant from us,  ten times more as an instance, the peak magnitude of the flare and the quiescent magnitude would be 18.5 mag and $\sim28.0$ mag, respectively. On the one hand,
a stellar flare with such "bright" peak could be detectable by lots of one meter class telescopes during their day-scale cadence surveys. However, since the fast decay after the peak, it is hard to obtain a spectrum before the end of the event due to a possible long-time delay from the discovery to the follow-ups after its successful identification. 
On the other hand, at the quiescent state, the counterpart could not be detectable for most of telescopes. 

Overall, the characteristics of a bright peak, a fast decay and a very faint counterpart at the quiescent state would result in an incorrect identification by 
treating a stellar flare as the best candidates of the cosmic event with a bad
localization.
Highly intense monitor of millions of stars for their activities would be helpful to flag these sources, and a fast spectrum even with a low-resolution at the early phase after an event could provide a very essential clue on its nature.

\section*{Acknowledgements}
The authors would like to thank the anonymous referees for their helpful comments and suggestions.
This study is supported from the National Natural Science Foundation of China (Grant No. 11973055, U1938201, U1831207, U1931133,12133003) and partially supported by the Strategic Pioneer
Program on Space Science, Chinese Academy of Sciences, grant
Nos. XDA15052600 and XDA15016500. 
JW is supported by the National Natural Science Foundation of China under grants 12173009 and by the Natural Science
Foundation of Guangxi (2020GXNSFDA238018).
YGY is supported by the National Natural Science Foundation of China under grants 11873003. 
We acknowledge the support of the staff of the Xinglong 2.16m telescope. This work was partially supported by the Open Project Program of the Key Laboratory of Optical Astronomy, National Astronomical Observatories, Chinese Academy of Sciences.
This work has made use of data from the European Space Agency (ESA) mission Gaia (https://www.cosmos.esa.int/gaia), processed by the Gaia Data Processing and Analysis Consortium (DPAC, https://www.cosmos.esa.int/web/gaia/dpac/consortium). Funding for the DPAC has been provided by national institutions, in particular the institutions participating in the Gaia Multilateral Agreement. 
This research has made use of the VizieR catalogue access tool, CDS, Strasbourg, France (DOI: 10.26093/cds/vizier). The original description of the VizieR service was published in A\&AS 143, 23

%

\vspace{5mm}


\section*{DATA AVAILABILITY}
The datasets generated during and/or analysed during the current study are available from the corresponding author on reasonable request.


%








\appendix


\bsp	
\label{lastpage}
\end{CJK}
\end{document}